# Single-Pixel Imaging with Neutrons

Yu-Hang He[1,2,†], Yi-Yi Huang[1,2,†], Zhi-Rong Zeng[3,4,†], Yi-Fei Li[1,2], Jun-Hao Tan[1,2], Li-Ming Chen[5,*], Ling-An Wu[1,2,*], Ming-Fei Li[6], Bao-Gang Quan[1,2], Song-Lin Wang[3,4] and Tian-Jiao Liang[3,4,*]

[1]*Institute of Physics, Chinese Academy of Sciences, Beijing 100191, China*
[2]*University of Chinese Academy of Sciences, Beijing 100049, China*
[3]*Institute of High Energy Physics, Chinese Academy of Sciences, Beijing 100049, China*
[4]*Songshan Lake Material Laboratory, Guangdong 523808, China*
[5]*Shanghai Jiao Tong University, Shanghai 200240, China*
[6]*Beijing Institute of Aerospace Control Devices, Beijing 100039, China*

† *Contributed equally to this work as co-first authors.*
\* *Corresponding author. Email: wula@iphy.ac.cn; lmchen@iphy.ac.cn; tjliang@ihep.ac.cn*

Neutron imaging is an invaluable noninvasive technique for exploring new science and assisting industrial manufacture. However, state-of-the-art neutron facilities are extremely expensive and inconvenient to access, while the flux of portable neutron sources is not strong enough to form even a static image within an acceptable time frame. It is hard to obtain images with both high spatial resolution and energy resolution together. Here, based on classical amplitude modulation, we demonstrate single-pixel imaging with neutrons with specially designed masks and, further, obtain energy-selective images with a spallation neutron source. Images of real complex objects with 100 μm spatial resolution and 10 μs time resolution (corresponding to 0.4% at 1 Å) have been obtained using a $^3$He single-pixel detector. Even when the neutron counts in the detector plane were lowered to 1000 per modulation pattern on average, a clear image was still obtained. The experimental setup is simple, inexpensive and easy to operate, thus our scheme points to a new path for neutron imaging, especially for portable radioactive neutron sources of low intensity, which should be of great benefit for diagnostic analysis in biology, materials science, and industrial processes.

Different from electrons and photons, neutrons are charge-free particles with magnetic moment which can easily penetrate metals but are very sensitive to the light elements, so can be used to complement x-rays. Neutron radiography is especially powerful in fields such as target diagnosis in inertial confinement fusion [1], inspection of batteries during industrial production [2], and magnetic structures analysis [3], where it is already a standard non-destructive tool [4]. In a typical high resolution neutron imaging system, a scintillation detector is always adopted. The neutrons deposit their energy in the scintillator to be converted into visible light which is then registered by a conventional high resolution charge-coupled detector (CCD). For better resolution, usually a thinner scintillator is desirable as it can produce light emission with a smaller point spread function, but this will be at the expense of detection efficiency so a tradeoff is necessary. Energy-selective imaging via time–of-flight (TOF) measurements [5] is even possible with spallation sources, which emit pulses of neutrons, in which case the energy resolution is mainly determined by the flux of the source and time response of the detector. When better energy resolution is required, the integration time of the detector must be shorter, which requires higher neutron flux. Basically, to form a clear neutron image, even for a common 2-dimensional (2-D) static image the radiation flux must be strong enough; this is generally possible with reactors or spallation sources but difficult to achieve with portable radioactive neutron sources. The former are expensive facilities that are hard to upgrade, thus it is extremely challenging to obtain images of both high spatial and energy resolutions with low flux illumination if only ordinary position-sensitive detectors are available. A possible solution is to perform second-order intensity correlation ghost imaging with a single-pixel (bucket) spectroscopic detector. Ghost imaging (GI) has been successfully performed with visible light [6], terahertz waves [7], microwaves [8], x-rays [9–12] and even particles—atoms [13] and electrons [14]. However, it is difficult to realize GI with neutrons because of the lack of suitable modulation devices and the inconvenience of having to go to a distant neutron source facility.

Ghost imaging was first realized with photon pairs produced by spontaneous parametric downconversion [15]. For a long time it was considered a specific feature of quantum light, until it was demonstrated using classical pseudothermal [16] and later true thermal light [17]. In this scheme a beamsplitter divides a spatially chaotic beam into two, and sends one to illuminate a target and then be collected by a bucket detector (so-called because it should collect all the transmitted light from the object), while the other travels through free space to a spatially resolving reference detector. Through second-order correlation of the 2-D data of the latter with the bucket intensities and after averaging over many frames, an image of the object can be retrieved. Later, computational GI [18,19] opened up a new avenue in which the high resolution reference detector is replaced by a spatial light modulator such as a digital micromirror device. It should be noted that so-called single-pixel imaging [20,21], based also on second-order intensity correlation, was developing in parallel; the main difference being that the non-spatially resolving bucket detector was called a single-pixel detector. Ghost imaging has now seen rapid development at many wavelengths in various applications, including super resolution imaging [22], remote sensing [23], hyperspectral imaging [24], and so forth. However, neutrons are fermions and may behave quite differently from bosonic photons in ghost imaging [25,26]. Although some theoretical work and experimental proposals about Fourier-transform ghost imaging have been written based on the Fermi-Dirac statistics and anti-bunching behavior of fermionic fields [27], prior experience suggests that single-pxel imaging with neutrons (SPIN) should be realizable in a classical way. For this, it is necessary to use some special modulation device to generate a measurable distribution of neutron speckles. Since their main interaction is with the nuclei of a material, neutrons can either be absorbed or scattered during propagation through a medium. To fabricate a neutron intensity modulation mask, a large absorption cross-section is obviously best. Although one can prerecord a series of randomly patterned modulation masks with a scintillator and high resolution pixelated detector then put the object in place and measure the bucket detector intensity at each precisely calibrated position, this is very time consuming and of low efficiency; moreover, the resolution is restricted by the features of the scintillator and the pixel size of its CCD.

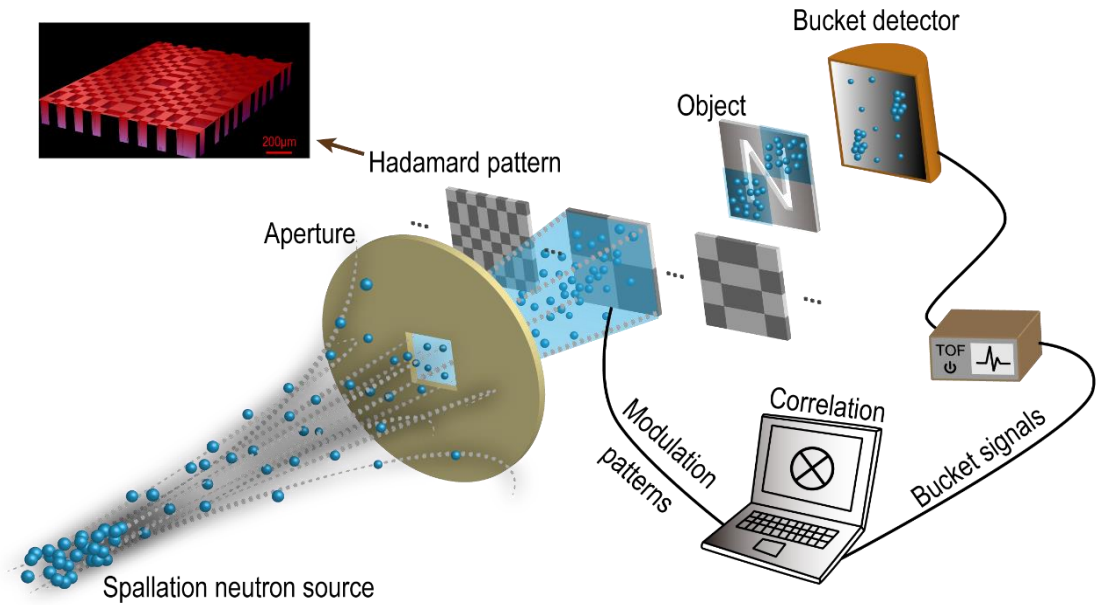

**Fig. 1.** (Color online) Experimental scheme of SPIN. The inset is a typical modulation pattern recorded by a 3-D optical microscope.

In this paper we report a relatively simple SPIN scheme by which images of high spatial and energy resolution are obtained with only a single-pixel spectroscopic detector and a polychromatic spallation neutron source of poor spatial coherence. The experiment was carried out in Beamline No. 20 at the China Spallation Neutron Source in Guangdong province. For the spatial modulator, the scheme of computational ghost imaging that has been successfully exploited for both high-energy electromagnetic radiation and particles [12–14,28] was used to realize SPIN. The setup of our experiment is shown in Fig. 1. Polychromatic neutrons were emitted from the spallation source at a repetition rate of 25 Hz, and emerged from the end of a 20 mm diameter collimator at a rate of around $10^7/cm^2/s$. According to geometrical optics, for an imperfect point source the resolution of the imaging system is mainly limited by the spatial coherence of the source [29]. Therefore, to improve the collimation, a 4 mm thick, 1.6 mm square aperture made of cadmium was put 22.7 cm downstream. About 23.8 cm further, a 15×15 cm modulation mask was mounted on a precise motorized translation stage. The mask consisted of a set of 1024 Hadamard patterns in a 32×32 matrix, and was aligned into the neutron beam with the aid of a collinear laser guide beam. The patterns were projected in sequence one by one onto the target, which was placed as close as possible, about 1.2 cm away. The transmitted neutrons were finally collected by a bucket detector at a distance of 27.3 cm behind the object. For this single-pixel detector we employed an LND custom designed proportional counter system of 10×10 cm effective area filled with $^3$He gas. This detector was chosen on account of its high absorption cross section and detection efficiency for thermal neutrons, low sensitivity to gamma rays, and high signal-to-noise ratio. Thermal neutrons (1 - 10 Å) are detected through the $^3$He (n, p) $^3$H process with a reaction cross section of 5330 barns, which will decrease in inverse proportion to the neutron velocity, up to about 0.2 MeV. When a neutron enters the detector, it interacts with a $^3$He isotope to produce a 191 keV proton and a 573 keV tritium isotope, which then ionize the surrounding proportional counter gas atoms to create more charges in an avalanche-like multiplication process, thus dissipating their combined energy of 764 keV. All the charges are collected by the detector which emits an output pulse proportional to 764 keV. It should be noted that although the detector is quite insensitive to gamma rays, due to the high intensity background of fast neutrons and gamma rays of the neutron beamline, some gamma ray noise would still be included in the detector output. These could be filtered out by setting a time-of-arrival threshold since gamma rays travel faster than neutrons, but unfortunately at the time of our experiment the spallation source was still in trial operation, and had not yet been installed with a $T_0$ time chopper. Each mask pattern was exposed for a certain length of time, depending on the sample and number of counts recorded, which will be explained later. After completion of all the 1024 matrix exposures, the image was retrieved by calculating the second-order correlation of each bucket signal with its corresponding matrix pattern, then averaging over the total.

The modulation patterns that interact with the object in the (x, y) plane may be denoted by $I_m(x,y)$, where $m = 1, 2, \ldots, M$. The 1-D intensity signal acquired by the bucket detector can then be written as

$$s_m = \int C\big(T_{obj}(x,y,\lambda_i), T_{mask}(x,y,\lambda_i), x, y, m\big) \phi_0(x,y) I_m(x,y) T_{obj}(x,y,\lambda_i) dx dy, \quad (1)$$

where $C(T_{obj}, T_{mask}, x, y, m)$ is a buildup factor [30] that expresses the dependence of the scattered flux on the image formation process, $T_{obj}(x,y,\lambda_i), T_{mask}(x,y,\lambda_i)$ are the transmittance functions of the object and mask at the incident neutron wavelength $\lambda_i$, respectively, and $\phi_0(x,y)$ is the spatial distribution of the neutrons arriving at the plane of the object (when no mask is present). The pattern distribution $I_m(x,y)$ determines the modulated intensity $T_{mask}(x,y,\lambda_i)$ trasnmitted through the mask. For simplicity, we assume that our sample is only slightly scattering, so we can take $C(T_{obj}, T_{mask}, x, y, m) \approx 1$. This is reasonable because the absorption is much stronger than scattering. The image is then reconstructed from the correlation of the intensity fluctuations with the illumination speckle patterns, but first it is necessary to normalize using the neutron background distribution, so we obtain

$$\tilde{G}(x,y) = \langle \Delta s_m \Delta I_m \rangle / \phi_0(x,y), \quad (2)$$

where $\langle \cdot \rangle$ denotes an ensemble average over $M$ measurements and $\langle \Delta s_m \rangle = s_m - \langle s_m \rangle$, $\langle \Delta I_m \rangle = I_m - \langle I_m \rangle$. To recover the images we also employed an effective convolutional neural network (CNN) algorithm [31] which, as will be shown below, greatly enhanced the image quality.

One of the trickiest part of our experiment was to fabricate a suitable modulation mask. First, gadolinium oxide ($Gd_2O_3$) was selected as the modulation material as it has a very large resonance and absorption cross section for thermal neutrons, as well as stable chemical properties. A silicon substrate was etched with 1024 Hadamard patterns, each containing 32×32 pixels, and each pixel about 100 μm square in size and 300±30 μm deep. A typical pattern is shown in the inset of Fig. 1, recorded by a 3D optical microscope (Bruker Contour GT), where we can see that the edge of each pixel is very steep. $Gd_2O_3$ powder was then compacted into the silicon mould; this was performed manually so it was not possible to fill in all the hollow spaces perfectly, but since the $Gd_2O_3$ particles were of nanometer size, about three orders of magnitude smaller than the size of the etched pits, the job was reasonably well accomplished. Finally we obtain a couple of modulation masks. By direct neutron imaging with a Fuji ND imaging plate, we measured the modulation depth ratio to be about 60%, which is sufficient for SPIN.

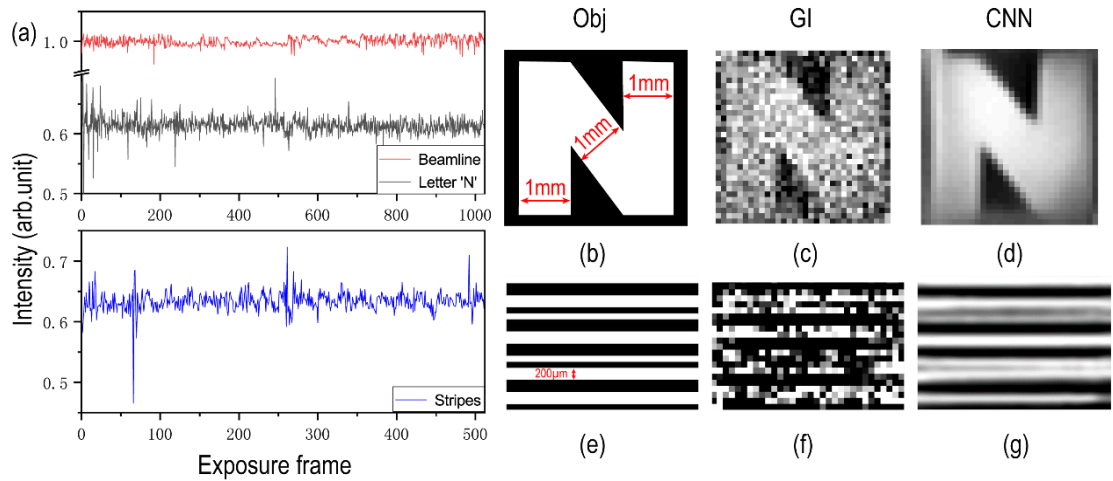

**Fig. 2.** (Color online) (a) Bucket intensities recorded for each exposure frame; from top to bottom: beamline fluctuations, for the letter 'N' (top), and for the stripes. (b) – (d) Object 'N' and images retrieved by conventional GI and CNN, respectively, from 1024 exposures. (e) – (g) Striped object and images retrieved by conventional GI and CNN, respectively, from 512 exposures.

In computational ghost imaging [18] the resolution is determined by the pixel size of the modulation pattern, in our case 100 μm. To test the practicability of our scheme, two objects were taken. The first was a letter 'N' hollowed out of a 4 mm thick sheet of cadmium, the width of each line stroke being 1 mm, as shown as Fig. 2(b). The second object, shown in Fig. 2(e), consisted of a set of stripes with widths of 100 μm and 200 μm, made similarly to the mask by etching grooves in a Si substrate and filling with $Gd_2O_3$ powder.

To obtain a sufficiently large bucket signal we set an exposure time of 40 s and and 30 s, respectively, for the two objects so that both produced roughly equal counts of ~9000 per matrix pattern. For the letter N the total number of exposures was 1024, while for the stripes 512 was sufficient. However, the measured counts are not the real effective bucket signal as it includes background noise which arise primarily from two sources and should be accounted for. First, as mentioned above, the depth of the mask structures were not perfectly uniform due to fabrication difficulties. This created a nonuniformity of about 10%, as checked by a stylus profiler (DektakXT Stylus Profiler, Bruker) and 3-D optical microscope (Bruker Contour GT), so to compensate for this a series of bucket detector measurements was initially recorded without the object in place, to be used in normalization. Second, fluctuations of the protons that produce the neutrons in the spallation source also generate noise. This is beyond our control, but assuming that the neutron production rate is proportional to that of the protons in the spallation source, the noise can be removed by normalization according to the actual real time data of the proton intensity, which was monitored and supplied by Beamline No. 20; this is shown in the topmost plot of Fig. 2(a). The twice normalized bucket signals of the letter N and the stripes are shown in the central and bottom plots of Fig. 2(a), respectively.

To begin with, the energy spectrum of the neutrons that we integrated over ranged from 1 to 10 Å. The images retrieved by conventional GI and also by a multi-level wavelet CNN algorithm are shown in Figs. 2(c), (d) and (f), (g) for 'N' and the stripes, respectively. It can be seen that the CNN results are markedly better, giving images with clear contours. Moreover, from Fig. 2(g) we see that a resolution of ~100 μm has been achieved.

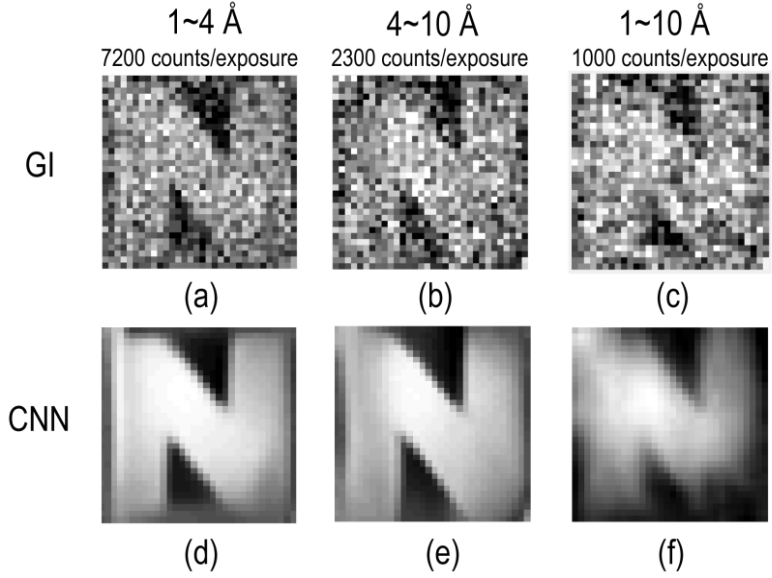

**Fig. 3.** Images retrieved after 1024 exposures for an energy spectrum ranging from (a), (d) 1—4 Å, 7200 counts/exposure; (b), (e) 4—10 Å, 2300 counts/ exposure; (c), (f) 1—10 Å, 1000 counts/ exposure. Top row: images retrieved by conventional GI; bottom row: images retrieved by CNN.

Since the velocity of the thermal neutrons depends on their energy, it is straightforward to differentiate their energy/wavelengths from TOF measurements with a single-pixel detector, and thus realize energy-selective SPIN. To demonstrate this, two separate series of bucket signals covering a spectral range of 1 to 4 Å and 4 to 10 Å were extracted, with a corresponding count rate of about 7200 and 2300 per matrix pattern exposure, respectively. The images from the two spectral ranges are shown in Figs. 3(a), (d) and (b), (e), respectively. In an energy selective neutron spectrometer, the spectral resolution is given by

$$\Delta\lambda = h\Delta t/mL, \qquad (3)$$

where $m$ is the mass of the neutron, $h$ is Planck's constant, $L$ is the distance from the neutron source to the detector and in our case is 9.7 m, and $\Delta t = 10$ μs is the time resolution of our detector. The spectral resolution is thus $\Delta\lambda/\lambda = 0.4\%$ at 1 Å. Of course, to achieve this resolution there must be sufficient counts in the spectral bandwidth of interest, which would require higher beam intensities or higher sensitivity detectors. In this case, the advantage of SPIN becomes apparent, because a simple $^3$He counter can register a much lower neutron flux than an array detector.

To test the feasibility of our scheme with low beam intensities, we reduced the exposure time down to 4 s for each of the 1024 mask patterns and obtained bucket signals with average counts of only 1000 through the whole spectrum of 1 to 10 Å; the results are shown in Figs. 3(c) and (f). The images in the upper row of Fig. 3 were obtained by conventional second-order correlation GI and those in the lower row by CNN; it is evident that the latter are much better than the former. Even with only 1000 counts per exposure the letter N in Fig. (f) is quite legible. Since cadmium does not have any resonant absorption feature within the spectrum of interest, Figs. 3(a) to (c) are quite similar, as also Figs. (d) to (f); there is only a slight difference in brightness due to a small difference in absorption. However, the experiments demonstrate the capabilities of our scheme for spectroscopic SPIN.

In conclusion, based on classic intensity modulation, we have successfully demonstrated low flux energy-selective spectral SPIN of real objects containing complex structures with 100 μm spatial resolution and a 10 μs time resolution (corresponding to 0.4% at 1 Å) with the $^3$He detector used in our experiment. A real bucket detector was used which greatly lowered the neutron flux required. The experimental setup is very simple, inexpensive and easy to operate. It can be quickly installed even in a beamline that was not originally designed for imaging but contains a neutron diffraction spectrometer or triaxial spectrometer; some necessary auxiliary imaging experiments could be performed at the same time. The source does not have to be coherent, and high resolution can be achieved even with portable low flux radioactive decay neutron sources. As a pioneering proof-of-principle experiment we did not explore the limits of our scheme, but clearly the energy resolution could be further improved if a detector with higher time resolution were used, and the sampling time reduced through optimized design of the masks and more efficient algorithms. There is indeed great potential for the applications of SPIN in basic research in biology and materials science, as well as in industrial product development such as diagnosis of batteries, fuels, and so forth.

**Acknowledgement**
This work was supported by the National Key R&D Program of China (2017YFA0403301, 2017YFB0503301, 2018YFB0504302), National Natural Science Foundation of China (11721404, 11805266), the Key Program of CAS (XDB17030500, XDB16010200) and the Science Challenge Project (TZ2018005).